\documentstyle[mymulticol,pra,aps,amsfonts]{revtex}

\newcommand{\cR}{{\cal R}}

\newcommand{\beq}{\begin{equation}}
\newcommand{\eeq}{\end{equation}}
\newcommand{\beqy}{\begin{eqnarray}}
\newcommand{\eeqy}{\end{eqnarray}}

\newenvironment{Definition*}{{\bf Definition}}{}

\makeatletter
\def\@beginTheorem#1#2{\trivlist \item[\hskip \labelsep{\bf #1\ #2}]}
\def\@opargbegintheorem#1#2#3{ \trivlist
      \item[\hskip \labelsep{\bf #1\ #2\ (#3)}]}
\makeatother

\makeatletter
\def\@beginLemma#1#2{\trivlist \item[\hskip \labelsep{\bf #1\ #2}]}
\def\@opargbeginLemma#1#2#3{ \trivlist
      \item[\hski
 Hence we have the same statements
about the increase of the supports for increasing depth, where
the local transformations are not counted for the depth.
p \labelsep{\bf #1\ #2\ (#3)}]}
\makeatother

\makeatletter
\def\@beginDefinition#1#2{\trivlist \item[\hskip \labelsep{\bf #1\ #2}]}
\def\@opargbeginDefinition#1#2#3{ \trivlist
      \item[\hskip \labelsep{\bf #1\ #2\ (#3)}]}
\makeatother

\makeatletter
\def\@beginCorollary#1#2{\trivlist \item[\hskip \labelsep{\bf #1\ #2}]}
\def\@opargbeginCorollary#1#2#3{ \trivlist
      \item[\hskip \labelsep{\bf #1\ #2\ (#3)}]}
\makeatother

\makeatletter
\def\@beginExample#1#2{\trivlist \item[\hskip \labelsep{\bf #1\ #2}]}
\def\@opargbeginExample#1#2#3{ \trivlist
      \item[\hskip \labelsep{\bf #1\ #2\ (#3)}]}
\makeatother

\def\C{{\mathbb{C}}}

\def\R{{\mathbb{R}}}

\newcommand{\cH}{{\cal H}}

\title{Quantum algorithm for measuring the energy   of  $n$ qubits
\newline with unknown pair-interactions}

\author{Dominik Janzing\thanks{Electronic address: janzing@ira.uka.de}}
\address{Institut f\"ur Algorithmen und Kognitive Systeme (Prof. Th. Beth),
Arbeitsgruppe Quantum Computing,\\ Am Fasanengarten 3a,
    D--76\,131 Karlsruhe, Germany}

\begin{document}
\maketitle

\begin{abstract}
The well-known algorithm for quantum phase estimation
requires that the considered unitary 
is available as a conditional transformation
depending on the quantum state of an ancilla register.
We present an algorithm converting an unknown $n$-qubit 
pair-interaction Hamiltonian into a conditional one such that
standard phase estimation can be applied 
to measure the energy. Our essential assumption 
is that
the considered system can be brought into interaction with a quantum computer.
For large $n$ the algorithm could still be applicable
for estimating the density of energy states and might 
therefore be useful
for finding energy gaps in solid states.
\end{abstract}

\begin{multicols}{2}

\section{Introduction}

Finding the energy spectrum of a given Hamiltonian
is an important task in physics
since it determines  the dynamical and thermodynamical behavior of a 
quantum system.
In solid-states physics, for
instance, the spectrum, in particular its gaps, are relevant for
the transport properties and in the famous  BCS-theory \cite{AsMe}
the energy gap between the ground state and the first excited state
is decisive for understanding super-conductivity.

Therefore a lot of efforts has been done on calculating the spectrum of 
many-particle Hamiltonians. Some models are known, where 
the Hamiltonian can explicitly be diagonalized   
\cite{Ba} and 
advanced mathematical tools like 
non-commutative geometry could be shown to be useful for
finding  energy gaps in  perfect crystals or those  with 
defects and quasi-periodicity \cite{Bel}.
Nevertheless, the diagonalization of generic many-particle Hamiltonians
is computationally hard and even restricted information about its spectrum
is difficult to obtain, since the dimension of the Hilbert space grows 
exponentially with the number of particles.
Here we discuss how to obtain information about the spectrum
of a Hamiltonian by a
quantum computer. Apart from the fact that the algorithm is able to find energy
gaps in many-particle systems efficiently, it has a property which is
impossible in principle without processing quantum information:
we can gain information about the spectrum of a physical  system with an 
{\it unknown} 
Hamiltonian provided that there is an interface between the quantum
 register and the considered system for exchanging quantum information\footnote{Of course information about the spectrum of a system's Hamiltonian can be 
obtained by sophisticated experiments in principle without using 
our algorithm. But performing sophisticated quantum measurements {\it is} 
quantum information processing.}. If such an interface does not exist
the algorithm can be used if the Hamiltonian is known and the quantum computer
is able to {\it simulate} the corresponding unitary evolution,
i.e., if efficient implementations of $\exp(-iHt)$ for each $t>0$ exist.
Then the algorithm can use those implementations as 
black-box subroutines\footnote{The problem of simulating pair-interaction
Hamiltonians
by a given one has recently been discussed in \cite{DNB,WJB,JWB,BCL,Leu}.}.

We restrict our attention to an unknown Hamiltonian $H$ 
of the following form. Let $H$ 
act on the Hilbert space $\cH:=(\C^2)^{\otimes n}$, which we shall refer to
as the  {\it target register}.
We assume $H$ to consist of 1-qubit terms and pair-interactions between
the $n$ two-level systems, i.e.
\begin{equation}\label{JForm}
H:= \sum_{j \leq n, \alpha} r^j_\alpha \sigma_\alpha^j +  
\sum_{k<l\leq n,\alpha,\beta}
J_{k,l,\alpha,\beta} \sigma_\alpha^k \sigma_\beta^l,
\end{equation}
where $\alpha =x,y,z$ and $\sigma^k_\alpha$ is the Pauli matrix 
$\sigma_\alpha$ acting on qubit $k$.  
The $3n\times 3n$-matrix $J$ and the $3n$-vector $r$ specify the interaction
uniquely.

First we should discuss why the problem can not be solved by a
simple application of the
well-known quantum algorithm for
phase estimation \cite{NC,TMi}.
The reason is that
the algorithm for estimating eigenvalues of a unitary $u$
does {\it not} work with black-box queries of $u$, it  relies on
the {\it quantum} transformation 
\[
\tilde{u}:=u \otimes  |1\rangle \langle 1|
+ 1 \otimes |0\rangle \langle 0| 
\]
implementing $u$ if and only if
an ancilla qubit is in the state $|1\rangle$, i.e.,
the ancilla qubit controls the implementation of $u$.
This causes severe problems if $u:=\exp(-iHt)$ is the natural time evolution, 
since phase estimation  would require a system with Hamiltonian
\[
H \otimes |1\rangle \langle 1|,
\]
i.e., the time evolution should be switched on and off by the 
ancilla's state.
First of all it is not clear how to  `switch off' the natural
dynamics of a system, for example the interaction
between nuclear spins in a real molecule.
However, this problem can be solved if certain approximations are allowed
as it is the case in standard
decoupling techniques in Nuclear Magnetic Resonance \cite{SMa}.
Assume that local transformations of the form
\begin{equation}\label{vau}
v:=v_1\otimes v_2 \otimes \dots \otimes v_n
\end{equation}
can be applied arbitrarily fast (`fast control limit').
If $v$ is applied before the system evolves according to its Hamiltonian
$H$ and $v^\dagger$ is applied afterwards, then the
system evolves as if it was subjected to the 
conjugated Hamiltonian $vHv^\dagger$.
Concatenations of such `conjugated evolutions'
for small time intervals make the system approximatively evolving
as if it was subjected to the `average Hamiltonian' which is
given by the convex combination of the different conjugated
Hamiltonians \cite{WJB,BCL}.
This technique can be applied for  `switching off'
Hamiltonians \cite{SMa}.
At first sight it might seem as if this technique could be applied
to control $H$ by the ancilla's state. But this would require
to substitute the sequence of decoupling local transformations 
by conditional transformations of the form
\[
\tilde{v}:=v \otimes |0\rangle \langle 0|  +  1\otimes |1\rangle \langle 1| .
\]
with $v$ as eq. (\ref{vau}).
The assumption that  $\tilde{v}$ 
could be implemented arbitrarily fast is much less justified
than the usual fast control limit, since such a transformation
refers to an interaction between ancilla and the target qubits
and we have no reason to assume that this interaction
is considerably stronger than the interaction which we want to switch off.

In other words, there is {\it no obvious} way to convert $H$ into
$ H \otimes |1\rangle \langle 1|$.
Below we present a scheme converting $H$ into $H \otimes \sigma_z$, which has  
essentially the same effect.
Our assumptions are the following.

\begin{enumerate}
\item
The target register $\cH$ 
can be brought 
into interaction with a quantum computers register $\cR$. This interaction acts
on the same time scale as $H$. 
The interaction between $\cH$ and $\cR$ can be switched off without
disturbing $H$.

\item  
During that time period where $\cH$ interacts with $\cR$ the Hamiltonian
evolution $H$ can be switched off.
This decoupling is controlled by classical signals
and not by a state of a quantum register.
\end{enumerate}

In order to illustrate this assumptions we 
consider a molecule with $k:=n+m$ nuclear spins.
Assume that the interactions between the spin pairs
$(j,l)$ for  $j,l \leq n$ are
unknown and the interactions between all the other pairs are known.
Given an arbitrary subset $M\subset \{1,2,\dots, n+m\}$, the
following procedure can be applied for switching off all those 
terms in the Hamiltonians concerning qubits in the set $M$.
For each qubit in  $M$
choose vectors $x_1,\dots,x_l$ of equal dimension $d$
with the following property:
The entries are  $1,x,y,z$ and for each pair $x_i,x_j$ of vectors 
each pair $(\alpha, \beta)$ with $\alpha,\beta=1,x,y,z$ appear
equally often in the list $(x_i^r,x_j^r)_{r\leq d}$ if $x_i^r$ is the $r^{th}$ entry of the vector $x_i$. 
The vectors $x_1,\dots, x_l$ are said to form an {\it orthogonal array} 
\cite{SMa}.
If $x^r_j=\alpha$ than the time evolution of spin $j$ is conjugated 
by the unitary transformation $\sigma_\alpha$ during the $r^{th}$ time period,
where we have used the convention that  $\sigma_1$ is the identity map.
If the time periods are small, the resulting time evolution
is approximatively 
the identity for all the spins in $M$ and
those terms of $H$ which do not involve spins in $M$ are unchanged.

Then we consider the first $n$ spins as the target register 
with the unknown Hamiltonian $H$ and the other spins form the
quantum computer's register $\cR$.
The total Hamiltonian will be of the same form as eq. (\ref{JForm}) 
characterized by
a $3(n+m)$--vector $r$ and a $3(n+m)\times 3(n+m)$--matrix $J$.

In order to switch off the interaction between
$\cH$ and $\cR$ we choose $M:=\{n+1,\dots,n+m\}$. Then 
the unknown  Hamiltonian $H$ on the spins $1,\dots,n$  is remaining.

If the Hamiltonian on the target register should be switched off in such a way
that the interaction between one specific qubit $j\leq n$
in the target register and one specific qubit $n+l$ in the ancilla register
remains, one has to take $M$ as the complement of the set $\{j,n+l\}$.
Note that this does not remove the 1-qubit terms $\sigma_\alpha^{n+l}$
and $\sigma_\alpha^j$. This is important since
 $\sigma_\alpha^j$ is unknown and is therefore
disturbing if we want to implement a definite 2-qubit transformation 
on the qubit pair $(j,n+l)$. Therefore we have to switch it off.
This can be done as follow. The interaction between $j$ and $n+l$ is
given as 
\[
\sum_{\alpha,\beta} g_{\alpha,\beta} \sigma_\alpha^j\sigma_\beta^{n+l} \,.
\]
Choose a specific pair $\alpha,\beta$ such that 
$g_{\alpha, \beta}\neq 0$.
Cancel all the terms $\sigma^j_{\tilde{\alpha}} \sigma^{n+l}_{\tilde{\beta}}$
 with $(\tilde{\alpha},\tilde{\beta})\neq 
(\alpha, \beta)$ by conjugation of the evolution with
the 4 unitaries $1, \sigma_\alpha^j, 
\sigma_\beta^{n+l}, \sigma_\alpha^j \sigma_\beta^{n+l}$ on $4$ time periods
of equal length.
The result is that the term 
\[
r^j_\alpha \sigma_\alpha^j + r^{n+l}_\beta\sigma_\beta^{n+l} +
g_{\alpha,\beta} \sigma_\alpha^j \sigma_\beta^{n+l}
\]
is remaining.
The 1-qubit terms can be cancelled by conjugating the  evolution with
$\sigma_{\alpha'}^j \sigma_{\beta'}^{n+l}$ for half of the time period
with $\alpha'\neq \alpha$ and $\beta'\neq \beta$.
Since this changes the sign of the operators $\sigma_\alpha^j$ 
(by the anti-commutation property of the Pauli-matrices) and
$\sigma_\beta^l$, the bilinear term $\sigma_\alpha^j\sigma_\beta^k$ 
is unchanged.

This shows that the assumptions 1.~ and 2.~ above are justified.
Note that it is not relevant which interaction between ancilla
qubits and register is available. If the true physical interaction
is 
\[
\sum_{\alpha,\beta} g_{\alpha,\beta} 
\sigma_\alpha^j \sigma_\beta^{l+n}
\]
then it can be converted into each other Hamiltonian
with coefficients $\tilde{g}_{\alpha,\beta}$ (see \cite{BCL}) in the sense
of the `average Hamiltonian' method.
Hence, if we write `switch on the interaction  $\sigma_x\otimes \sigma_z$' 
we do not assume the real physical Hamiltonian to be of this form, 
we assume only that the true Hamiltonian of the system
can be used for simulating the required term. This is the continuous analogue
of the usual way of describing algorithms by basic gates as `controlled not'
operations: it does not matter whether the quantum computer really has 
the `controlled not' as a basic operation but it should be capable
of `simulating' it by those  transformations which are really available.
Our way of describing the algorithm by `switching on and off interactions'
should therefore only be considered as a convenient 
{\it language} for continuous quantum algorithms.

\section{The algorithm}

First we consider only one ancilla qubit and describe how to convert
the Hamiltonian $H\otimes 1$ into the conditional Hamiltonian
$H\otimes \sigma_z$ controlled by the ancilla's state.

We 
can switch off all the coupling except of those on one
specific qubit pair $(j,k)$ by so-called selective decoupling
\cite{Leu}. Furthermore we can select those terms in the remaining Hamiltonian
which contain only $\sigma_\alpha^j$ and $\sigma^k_\beta$ for specific 
$\alpha$ and $\beta$.
The remaining term is given by
\[
H_{j,k,\alpha,\beta}:= r_\alpha^j \sigma_\alpha^j + r_\beta^k\sigma_\beta ^k +
J_{j,k,\alpha,\beta} \sigma^j_\alpha \sigma^k_\beta
\]
First we convert $H_{j,k,\alpha,\beta}$ into the term
\[
H'_{j,k,\alpha,\beta} :=\frac{2}{n-1}(r^j_\alpha \sigma_\alpha^k + r^k_\beta
\sigma_\beta^k)
+J_{j,k,\alpha,\beta} \sigma_\alpha^j \sigma_\beta^k . 
\]
This can be done by dividing the considered  small time interval $[0,\delta]$
into two intervals $[0,\delta(1/2 -1/(n-1))]$ and $[\delta(1/2 -1/(n-1)),\delta]$.
During the first interval the evolution according to $H_{j,k,\alpha,\beta}$
is conjugated by 
$\sigma_\alpha^j\sigma_\beta^k$. This reverses the sign of the 1-qubit terms
whereas the 2-qubit term is unchanged.
During the second interval the  evolution according to 
$H_{j,k,\alpha,\beta}$ is applied.

Now we can describe the conversion into a controlled
Hamiltonian. 
By identifying $x,y,z$ with the additive group $F_3:=\{0,1,2\}$ 
it can be explained as follows.

\begin{enumerate}
\item
Apply the  unitary transformation $u_j$ defined by
the equation
\[
u_j \sigma_\alpha^j u_j^\dagger = \sigma^j_{\alpha+1}.
\]

\item
Apply time evolution corresponding to $H'_{j,k,\alpha,\beta}$ 
for  a small time period $\epsilon$.

\item
Switch on the interaction between target and ancilla qubit given by
\[
H_j:=\sigma_{\alpha +1}^j \otimes \sigma_z 
\]
for the time $\epsilon$.

\item
Simulate evolution according to  $-H'_{j,k,\alpha,\beta}$ for 
the time $\epsilon$.
Inverting unknown Hamiltonians is described in
\cite{Leu}.
In our case 
the 
inversion subroutine can be reduced as follows.
Choose a large number $p$. Divide the time interval
$[0,3\epsilon]$ into $3p$ intervals of equal length.
During the  interval $l$ conjugate the evolution 
according to $H'_{j,k,\alpha,\beta}$ by
the unitary transformations $\sigma_{\alpha+1}^j, \sigma_{\beta+1}^k,
\sigma_{\alpha+1}^j \sigma_{\beta+1}^k$  if $l$ is $0,1,$ or $2$
mod 2, respectively. In the limit $p\to\infty$, we 
obtain  a perfect simulation of the evolution according to
$-H'_{k,l,\alpha,\beta}$.
 
\item
Switch on the interaction
\[
-H_j
\]
for the time $\epsilon$.

\item
Apply $u^\dagger_j$.

\item Repeat steps 1 to 6 with qubit $k$.

\end{enumerate}

Up to an error of order $\epsilon^3$  
this simulates the time evolution corresponding to
\[
i u_j [ H'_{j,k, \alpha,\beta} 
, H_j]  u^\dagger_j +
i u_k [ H'_{j,k, \alpha,\beta} 
, H_j]   u^\dagger_k
 =
2 H''_{j,k,\alpha, \beta} \otimes \sigma_z 
\]
for the  time $\epsilon^2$ with  the definition
\[
H''_{j,k,\alpha,\beta}:= \frac{1}{n-1}(r^j_\alpha \sigma_\alpha^j
+r^k \sigma^k_\beta) + J_{j,k,\alpha,\beta} \sigma_\alpha^j \sigma_\beta^k.
\]

Due to the equation
\[
\sum_{j<k\leq n, \alpha,\beta} H''_{j,k,\alpha, \beta} =H
\]
the concatenation of the above procedure  for all unordered pairs  
$(j,k)$ and all $\alpha, \beta$  simulates time evolution according to
\[
H\otimes\sigma_z.
\]

Note that the time required for obtaining the unitary  transformation
\[
\exp(-i H\otimes \sigma_z \delta)
\] 
goes to zero for $\delta \to 0$ but the time {\it overhead}  required
for this simulation depends on the desired accuracy since
the resulting evolution is only of second order a non-trivial one.
Note that $\sigma_\alpha^j \sigma_\beta^k$ can also be converted
into $\sigma_\alpha^j \sigma_\beta^k \otimes \sigma_z$  by
a conjugation of the evolution with
an appropriate 2-qubit gate acting on qubit $k$ and the ancilla qubit, since 
$\sigma_\beta^k\otimes 1$ and $\sigma_\beta^k\otimes \sigma_z$ have the 
same spectrum as operators on 2 qubits.
But this kind of converting the Hamiltonian into a conditional one does
not have the property that an infinitesimal time step of the algorithm
requires only infinitesimal time.
Whether this is a true disadvantage for the desired accuracies 
has to be checked by thorough numerics.

Now we can apply usual phase estimation procedure
to the conditional Hamiltonian evolution.

Now we initialize the ancilla register by a Hadamard transformation on
each qubit.
We implement the transformation
\[
\exp(-i(H\otimes \sigma_z^j  2^j\tau))
\]
for each ancilla qubit $j=0,\dots,m-1$.
The total running time of this implementation 
is $(2^m-1) \tau$ times the time overhead for simulating
$H\otimes\sigma_z$ with appropriate accuracy.

We choose $\tau$ in such a way that $2 \tau \Delta =\pi$ if 
$\Delta$ is an  upper bound for the difference
between greatest and smallest eigenvalue of $H$ given by prior knowledge.
For every known systems 
in many-particle physics, $\Delta$ is always of the order of $n$, since
energy per particle is a well-defined quantity in the thermodynamic limit
$n\to\infty$.
The size $m$ of the ancilla register has to be chosen in such a way that
$2^{-m}\tau $ is smaller than the desired accuracy of the energy 
measurement. It is convenient to consider the ancilla qubit $j$ as the
$j^{th}$ number of a binary digit. Then the initial state of $\cR$
can be written as
\[
\frac{1}{\sqrt{2^m}}\sum_{l < 2^m}|l\rangle .
\]
If the target register is in an eigenstate of $H$ with eigenvalue
$E$ we obtain (up to an irrelevant global phase) the ancilla state
\[
\frac{1}{\sqrt{2^m}}\sum_l e^{-il2E\tau } |l\rangle\, ,
\]
since $\exp(-i\sigma_z\tau)$ produces a relative  phase difference 
$\exp(-2i\tau)$
between the ancilla states $|0\rangle$ and $|1\rangle$.

By inverse Fourier transformation, one obtains the state
\[
\frac{1}{2^m} \sum_{k,l} e^{-i 2\pi kl/(2^m)}e^{i2E\tau} 
|k\rangle. 
\]
Measuring the ancilla register in the standard basis provides
good estimations for $E$ since the probability that the 
measured
result $k $ differs more than $e$
from $2^m  E \tau/\pi$ is less than $1/(2e -2)$\cite{NC}.
If $2^{-m}/ \tau$ is much smaller than the energy gaps of $H$, the
algorithm projects onto the eigenstates of $H$ \cite{TMi}.
For large $n$, we cannot expect this to be achievable since
the gaps decrease exponentially with $n$ in the generic case.
However, in this case the algorithm can be used for
estimating the {\it density of energy states}, since
the measured results will mostly be around those energy values where
the density is high, provided that the target register has been initialized
in the maximally mixed state.
Estimating the density of states is
an important task in solid-state physics \cite{Lon} and spectral
gaps can be detected by this method. Most interesting energy gaps
in solid states physics do not depend on the number of particles 
\cite{Zim,AsMe}.
They could be detected without increasing the size of the ancilla register.
But one should emphasize that only those gaps can be detected which are 
localized around {\it typical} energy values according to the 
initial density matrix. Therefore it might be more useful to start
with thermal equilibrium states in the target register in order to
find gaps around values which are typical for the desired 
temperature\footnote{Concerning the electrons of solid states, for instance,
one is often 
interested in the energy gaps near the Fermi level \cite{Zim}.}.

We should mention a difficulty which appears for large $n$ and how it could 
be overcome.
We assumed that each qubit in the target register can be 
brought into interaction
with each qubit of the ancilla register. Keeping in mind the example with
nuclear spins in a molecule, one should take into account
that the strength of the interactions is strongly decreasing
with the distance between the spins. Accordingly, the interaction time
between ancilla and target spin has to be increased for large distances
and the resulting running time might become unacceptable.
Therefore we emphasize that
the algorithm can be rewritten in such a way, that it is not necessary
to have an interaction between each qubit in $\cH$ 
and each qubit in $\cR$. Instead, one can transfer the information
qubit by qubit into a quantum register.
To speak more precisely,
one can realize the algorithm within the following setting 
{\it instead} of assumptions 1) and 2) above:

\begin{enumerate}

\item
Apart from $\cR$, 
there is a quantum register $\cH'$ with $n$ qubits
such that the `information exchange' 
\[
w: |\phi\rangle \otimes |\psi \rangle \leftrightarrow |\psi \rangle \otimes |\phi\rangle
\]
can be implemented at definite times in such a way that 
$\exp(-iHt)$ can be implemented on $\cH'$ by the following subroutine:
\begin{itemize}
\item
implement $w$.
\item 
wait the time $t$.
\item
implement $w$.
\end{itemize}

\item
The system $\cH'\otimes \cR$ has full capabilities of quantum computation.

\end{enumerate}

Then $\cH$ can be substituted by $\cH'$ if 
the instruction `wait the time $t$' 
is substituted by the subroutine above.
Due to the quantum computation capabilities on $\cH'\otimes \cR$
we can clearly implement any desired transformation of the form 
$ \exp(-i \sigma_\alpha\otimes\sigma_z t)$ for all $t\in \R$ 
between any arbitrary qubit pair. This shows once more that  instructions
of the form 
`switch on the interaction $\sigma_\alpha^j \otimes \sigma_\beta^{n+l}$'
as above should not be taken to literally.
Having full capabilities of quantum computation, we can clearly
implement unitary evolutions corresponding to the interaction
$\sigma_\alpha^j \otimes \sigma_\beta^{n+l}$. The best way of `simulating'
this interaction is hardware-dependent.

In order to estimate the running time of the algorithm
one should keep in mind that $n(n-1)/2$ interaction terms have to be coupled
to the ancilla register (The selection of specific
coupling does not produce any time overhead \cite{WJB,Leu,SMa}). 
The time for converting one interaction term into
a conditional one seems to be independent of $n$ at first sight. 
But here might come in a subtle $n$-dependence as well:
it is possible that 
the simulation of one term $H_{j,k,\alpha,\beta}\otimes \sigma_z$ has
to be more exact if one wants to find spectral gaps for large $n$.
Estimating the error in the eigenvalues of the total Hamiltonian 
resulting from an error in the simulation of each coupling 
might be a difficult task of perturbation theory \cite{Kato}.
But the error of the total Hamiltonian is only $n^2$ times
the error of the single coupling terms. Since we are interested in 
of gaps of the order 1, it should  be possible to 
reproduce them with polynomial time overhead.

The algorithm shows a nice application of `simulations of
Hamiltonians' \cite{WJB,BCL,Leu}: The algorithm can used to find eigenvalues
of Hamiltonians by simulating the corresponding unitary evolution
on a quantum computer.

Apart from its possible application in many-particle physics,
the algorithm shows a new aspect for understanding the
double-role of self-adjoint operators as observables
and generators of transformations, since we have constructed
a measurement procedure for $H$ by using the evolutions $\exp(-iHt)$
as black-box procedures\footnote{In an other setting, this has already been
noted in \cite{JZAB}}.
This gives a new scheme for measuring non-trivial joint observables. 

The reader might get confused about one little paradox:
we claimed to measure the actual energy value of the system
 despite the fact that only energy {\it differences} have physical relevance
(as long as general relativity is not involved). This paradox is resolved by
specifying more precisely what the algorithm does. 
We assumed the Hamiltonian $H$
to be traceless. In case we start with a non-traceless one, only the traceless
part will survive the conversion procedure
$H\otimes 1 \mapsto H\otimes\sigma_z$. Hence the algorithm measures
the difference between the 
actual energy and the trace of $H$, i.e., the
average of the energy values over all the states.
This seems again to be paradox: in case the system is in a definite 
eigenstate,
why can the ancilla register `feel' the average energy over all the other
states despite the fact that the system never deviates from its actual
eigenstate?  This paradox is only pretended by the `infinitesimal
language'. Of course the system {\it does} 
deviate a little bit from the actual
eigenstate {\it during} the algorithm. Note that
the procedure simulating  $H \otimes\sigma_z$ consists of 
a long sequence of  transformations (close to the identity) 
which  do not leave the eigenstates of $H$ invariant.

\end{multicols}

\section*{Acknowledgements}

Thanks to P. Wocjan for useful corrections.
Part of this work has been supported by the European project
Q-ACTA and the DFG-project `verlustarme Informationsverarbeitung'.


\begin{thebibliography}{10}

\bibitem{AsMe}
N.~Ashcroft and D.~Mermin.
\newblock {\em Solid state physics}.
\newblock Holt, Rinehart and Winston, New York, 1976.

\bibitem{Ba}
R.~Baxter.
\newblock {\em Exactly solved models in statistical mechanics}.
\newblock Academic Press, London, 1989.

\bibitem{Bel}
J.~Bellissard.
\newblock Gap labelling theorems for {Schr{\"o}dinger} operators.
\newblock In {\em From Number Theory to Physics}, 1992.

\bibitem{DNB}
J.~Dodd, M.~Nielsen, M.~Bremner, and R.~Thew.
\newblock Universal quantum computation and simulation using any entangling
  Hamiltonian and local unitaries.
\newblock {\em LANL-preprint quant-ph/0106064}, 2001.

\bibitem{WJB}
P.~Wocjan, D.~Janzing, and T.~Beth.
\newblock Simulating arbitrary pair-interactions by a given Hamiltonian:
  Graph-theoretical bounds on the time complexity.
\newblock {\em LANL-preprint quant-ph/0106077}, 2001.

\bibitem{JWB}
D.~Janzing, P.~Wocjan, and T.~Beth.
\newblock Complexity of inverting $n$-spin interactions: Arrow of time in
  quantum control.
\newblock {\em LANL-preprint quant-ph/0106085}, 2001.

\bibitem{BCL}
C.~Bennett, J.~Cirac, M.~Leifer, D.~Leung, N.~Linden, S.~Popescu, and G.~Vidal.
\newblock Optimal simulation of two-qubit Hamiltonians using general local
  operations.
\newblock {\em LANL-preprint quant-ph/0107035}, 2001.

\bibitem{Leu}
D.~Leung.
\newblock Simulation and reversal of n-qubit Hamiltonians using Hadamard
  matrices.
\newblock {\em LANL-preprint quant-ph/0107041}, 2001.

\bibitem{NC}
M.~Nielsen and I.~Chuang.
\newblock {\em Quantum Computation and Quantum Information}.
\newblock Cambridge University Press, 2000.


\bibitem{TMi}
B.~Travaglione and G.~Milburn.
\newblock Generation of eigenstates using the phase estimation algorithm.
\newblock {\em LANL-preprint quant-ph/0008053}, 2000.


\bibitem{SMa}
M.~Stollsteimer and G.~Mahler.
\newblock Suppression of arbitrary internal coupling in a quantum register.
\newblock {\em LANL-preprint quant-ph/0107059}.

\bibitem{Lon}
R.~Longini.
\newblock {\em Introductory quantum mechanics for the solid state}.
\newblock John Wiley \& Sons, New York, 1970.

\bibitem{Zim}
J.~Ziman.
\newblock {\em Priciples of the Theory of Solids}.
\newblock Cambridge University Press, 1972.

\bibitem{Kato}
T.~Kato.
\newblock {\em Perturbation theory for linear operators}.
\newblock Springer, Berlin, 19966.

\bibitem{JZAB}
D.~Janzing, R.~Zeier, F.~Armknecht, and T.~Beth.
\newblock Quantum control without access to the controlling interaction.
\newblock {\em LANL-preprint quant-ph/0103022}, 2001.

\end{thebibliography}
\end{document}